\documentstyle[epsfig,aps]{revtex}

\begin{document}

\draft

\def\lsim{\lower.5ex\hbox{$\; \buildrel < \over \sim \;$}}
\def\gsim{\lower.5ex\hbox{$\; \buildrel > \over \sim \;$}}


\title{Neutrino oscillations under gravity: mass independent oscillation}  
\author{Banibrata Mukhopadhyay\thanks{bm@cc.oulu.fi} \\
{\sl Astronomy Division, P.O.Box 3000, FIN-90014 University of Oulu, Finland} 
}

\maketitle
\baselineskip = 18 true pt
\vskip0.3cm
\setcounter{page}{1}

\def\ch{\lower-0.55ex\hbox{--}\kern-0.55em{\lower0.15ex\hbox{$h$}}}
\def\lh{\lower-0.55ex\hbox{--}\kern-0.55em{\lower0.15ex\hbox{$\lambda$}}}       
\def\n{\nonumber}

\begin{abstract}

I discuss the possibility of neutrino oscillation in presence of gravity. In this respect
I consider the propagation of neutrinos in the early phase of universe and around black holes.
It is seen that whether the rest masses of a neutrino and corresponding anti-neutrino are considered to be 
same or not due to space-time curvature effect non-zero oscillation probability between the neutrino and
anti-neutrino states comes out. Therefore I can conclude that under gravity neutrino oscillation takes 
place which is independent of their rest mass.
 
\end{abstract}

\vskip1.0cm
\vskip2cm



It is an established fact that the neutrinos have three different kinds of flavor.
Also the choice of these three flavors fits snugly into the standard model of 
particle physics, which tells that all the conventional matter is constructed 
from quarks and leptons. The neutrino flavors, namely, electron, muon and tau
neutrinos belong to the lepton class. However, the discrepancy in between 
various observed and expected number of neutrino flavors occur. This is because,
they transform from one type of flavor to another. Therefore, if a detector
is built and set to detect for one type of neutrinos, while that type of neutrinos
are coming, some of them are converting to another type which is unsuited to the
detector and thus results a shortfall. This anomalies have been found 
in the cases of solar neutrino, atmospheric neutrino and the neutrino experiments
at accelerators or reactors. Actually the neutrino flavor states consist superposition
of different mass eigenstates. Thus the flavor states transform from one to another
as different mass eigenstates evolve differently. This is called neutrino oscillation.
However, if CPT is violated, for a particular flavor the neutrino and anti-neutrino state
may transform from one to other as they evolve differently. In that case, oscillation
may take place between a neutrino and an anti-neutrino for the same flavor.

Out of several implications, study of neutrino has got great interest as neutrinos are the
important ingredient of the energy produced and removed from the centre of Sun. Also
in supernova, neutrinos carry off the largest amount energy of exploding star. 
Moreover, neutrinos are considered as a prime dark matter candidate. Though the mass
of the neutrinos is very less, because of their large number, density is very great
and would thus exert a very potent gravitational effect helping to determine the
rate and pattern of galaxy formation. Therefore they may have significant role 
in the large scale structure of the universe.

With all the above facts in background, naturally neutrino oscillation phenomena
are of great interest for particle physicists, astrophysicists, as well as 
cosmologist, as the number density of neutrino in universe is directly related to it.
Usually it is believed that the mass of a neutrino is the basic factor to 
attribute the oscillation. More precisely, it is directly related to the rest mass square
difference between two neutrino mass eigenstates. However, all these statements
are made when the neutrinos are considered according to Galilean relativity.
Gasperini \cite{gas} first pointed out that due to presence of gravitational field
different neutrino flavors could be affected differently which violates equivalence principle and
thus neutrino oscilations may take place even if they are massless or of degenerate mass.  
Subsequently the effect of strong gravitational field on neutrino oscillation was discussed \cite{hl}.
Later on the neutrino oscillation was studied considering the violation of equivalence 
principle with LSND data \cite{mansar} and was shown that the data can be explained by 
degenerate or massless neutrinos with flavor non-diagonal gravitational coupling. 
In a same line of thought, Coleman and Glashow \cite{cg} showed that in special relativistic regime, when the maximum 
velocity of different neutrinos may differ each other, oscillation may happen even if they
are considered massless. The goal of this present letter is to study of oscillation
phenomena when the space-time curvature has an important effect. Thus in this case, 
oscillation depends on the gravitational coupling strength. In past neutrino oscillations 
in curved space-time have been studied by various events \cite{all}. However, in this
present letter my message is something different. Recently, it has been 
shown that under gravity which brings the CPT violating interaction, neutrino and 
anti-neutrino split in their energy level and thus causes their asymmetry 
in number density \cite{ms,sm}. Here I will start from that result and apply for neutrino
oscillation business. If the neutrinos and corresponding anti-neutrinos 
split up in energy level, that creates a difference in their evolution phases,
which is directly related to the transition over their mass eigenstates.

It was argued earlier \cite{dva,ab} that in a weak gravitational field the flavor
oscillation is possible with the phase strength of oscillation probability is
proportional to the gravitomagnetic field. However, the system was chosen non-relativistic
and the corresponding Hamiltonian was Schr\"odinger-like constructed by hand. The possible
violation of local Lorentz invarience in Kaon sector of standard model was studied
with both CPT conserving and violating cases \cite{hms}. It was also
pointed out that \cite{dva2} in case of a CPT non-conserved situation the neutrino
and anti-neutrino acquire different mass and thus there is a possibility of oscillation
between themselves. In a same line of thought, it was shown \cite{baren} that in absence of 
neutrino decay, in the case of CPT non-conserving situation, a neutrino evolves differently than the 
anti-neutrino of same flavor and therefore the oscillation takes place between them. 
In a recent work \cite{ad}, it has been shown that a linear superposition of two opposite
helicity states can be described as the eigenstates of the de Sitter Casimirs, 
which actually give rise to the majorana states. It has been argued, due to the space-time effects neutrinos and
anti-neutrinos acquire different mass terms. In this present letter, all such scenarios 
have been studied in the more realistic situations when the neutrinos are propagating in curved space-times. 
I will start with Dirac equation in the generalized curved
space-time and all the interaction terms play in our calculation govern
inherently. When neutrinos which are class of fermion, propagate in the strong gravitational
field, e.g. close to a black hole or in the early universe era when space-time is non-flat, 
spin of a neutrino
couples with the spin-connection of space-time and generates a gravitational interaction
even if there is no other interaction present. As much the curvature effect of space-time 
is less, this gravitational interaction strength reduces too. 

Following \cite{ms}, the Dirac Lagrangian density for a neutrino under the gravity background in
locally flat coordinate system can be written as
\begin{eqnarray}
{\cal L}=det(e)\bar{\psi}\left[(i\gamma^a\partial_a-m)+\gamma^a\gamma^5 B_a\right]\psi
={\cal L}_f+{\cal L}_I,
\label{lagf}
\end{eqnarray}
where
\begin{eqnarray}
B^d=\epsilon^{abcd} e_{b\lambda}\left(\partial_a e^\lambda_c+\Gamma^\lambda_{\alpha\mu} 
e^\alpha_c e^\mu_a\right)\label{bd}.
\label{bd}
\end{eqnarray}
Our chosen unit is, $c=\ch=1$.
Clearly ${\cal L}_I$ may be a CPT and Lorentz violating interaction if the background curvature coupling,
$B_a$, is constant in local frame \cite{ms}.
If ${\cal L}_I$  changes sign under CPT transformation, it is violating CPT. More precisely, under 
CPT transformation, if the curvature coupling, $B_a$, does not flip sign then ${\cal L}_I$ violates CPT (as
in any case the associated axial-vector, $\bar{\psi}\gamma^a\gamma^5\psi$, changes sign under CPT). 
Therefore, it is the nature of background metric which determines whether $B_a$ is odd under CPT or not
and finally which dictates the overall nature of interaction.

The dispersion relation for the neutrino and anti-neutrino fields in presence of CPT violating gravitational
interaction can be given as \cite{ms}
\begin{eqnarray}
\nonumber
E_{\nu} &=&  \sqrt{|{\vec p}|^2 + 2 \left(B_0 p^0 + B_1 p^1 + B_2 p^2 + B_3 p^3 \right) + B_a B^a - m^2} \\
E_{\overline{\nu}} &=& \sqrt{|{\vec p}|^2 - 2 \left(B_0 p^0 + B_1 p^1 + B_2 p^2 + B_3 p^3 \right) + B_a B^a - m^2}.
\label{edis}
\end{eqnarray}
As $E_\nu\neq E_{\overline \nu}$ (due to presence of curvature which appears as 
gravitational four vector $B$) the neutrino 
oscillation phase depends on $(E_\nu - E_{\overline \nu})$, and a non-zero oscillation probability
between the neutrino and anti-neutrino comes out. 

Now following \cite{baren}, let me consider two distinct orthonormal eigenstates, one is $|E_\nu>$
for spin up neutrino type and other is
$|E_{\overline \nu}>$ for spin down anti-neutrino type for same flavor. Further introduce a set of neutrino mass
eigenstates at $t=0$ as
\begin{eqnarray}
|A>=cos\theta\, |E_\nu>+sin\theta\, |E_{\overline \nu}>,\hskip0.5cm 
|B>=-sin\theta\, |E_\nu>+cos\theta\, |E_{\overline \nu}>.
\label{fl2}    
\end{eqnarray}
Then the oscillation probability from a neutrino state $|A>$ at $t=0$ to an anti-neutrino state $|B>$ 
for the same flavor at a later time $t=t_1$ can be found as
\begin{eqnarray}
\nonumber
P_{ab}&=&\left|\left[-sin\theta\, <E_\nu|+cos\theta\, <E_{\overline \nu}|\right]\left[e^{-iE_\nu t_1}
cos\theta\,|E_\nu>+e^{-iE_{\overline \nu} t_1}sin\theta\,|E_{\overline \nu}>\right]\right|^2\\
&=&sin^22\theta\,sin^2\delta
\label{pab}
\end{eqnarray}
where, 
\begin{eqnarray}
\delta=\frac{(E_\nu-E_{\overline \nu})t_1}{2}.
\label{ph}
\end{eqnarray}
Now for simplicity, considering a special case of the space-time such that $B_0p^0>>{\vec B}.{\vec p}$ and 
as the neutrinos are highly relativistic, $|{\vec p}|^2$ is much larger compared to the other terms in dispersion relation. 
Therefore, $\delta$ reduces as
\begin{eqnarray}
\delta=B_0\,t_1.
\label{phf}
\end{eqnarray}
It is very interesting to note that although both the neutrino and anti-neutrino are 
chosen of same rest mass, due to the presence
of gravity a non-zero phase factor arises which finally produces a non-zero oscillation probability in two
mass states. Therefore, I can conclude that in gravity background, e.g. in the anisotropic phase of
early universe when space-time is non-flat or close to a rotating black hole, oscillations exist between 
the neutrino and anti-neutrino, whether their rest masses are same or different. Actually, due to the effect
of space-time curvature, the {\it effective mass} of a neutrino and an anti-neutrino differs (solely due to
the gravitational coupling appears as additive terms over $m^2$ in the dispersion relation) 
that finally brings a non-zero oscillation probability.  
Below I calculate the oscillation phase according to the specific metrics.

First I consider the anisotropic phase of axially symmetric early universe.
Then following \cite{bsc}, I choose a simplified version of the Bianchi II model given as
\begin{equation}
d s^2 = -dt^2+S(t)^2\, dx^2+R(t)^2\,[dy^2+f(y)^2\,dz^2]-S(t)^2\,h(y)\,[2dx-h(y)\,dz]\,dz
\label{mat}
\end{equation}
where $f(y)=y$ and $h(y)=-y^2/2$.
The corresponding orthogonal set of non-vanishing component of tetrads (vierbiens) can be written as
\begin{eqnarray}
\nonumber
&&e^0_t = 1,\,\,e^1_x=f(y)R(t)S(t)/\sqrt{f(y)^2R(t)^2+S(t)^2h(y)^2},\,\,e^2_y=R(t),\\
&&e^3_z=\sqrt{f(y)^2R(t)^2+S(t)^2h(y)^2},\,\,e^3_x=-S(t)^2h(y)/\sqrt{f(y)^2R(t)^2+S(t)^2h(y)^2}.
\label{tetnonv}
\end{eqnarray}
Therefore, from (\ref{bd}) I get
\begin{eqnarray}
\nonumber
B^0&=&\frac{4R^3S+3y^2R\,S^3-2y\,S^4}{8R^4+2y^2R^2S^2}, \hskip0.5cm B^1=0\\
B^2&=&\frac{(4y\,R^2-8R\,S-y^3S^2)(R\,S^\prime-R^\prime S)}{8R^4+2y^2R^2S^2}, \hskip0.5cm B^3=0.
\label{bian2}
\end{eqnarray}
If the scale factors of space-time are considered to be same in all directions, 
$S(t)=R(t)$, only $B^0$ will remain non-zero and is given as
\begin{eqnarray}
\nonumber
B^0=-B_0=\frac{3y^2-2y+4}{2y^2+8}.
\label{b0f}
\end{eqnarray} 
Therefore, from (\ref{phf}) neutrino oscillation phase at time $t_1$ in early universe reduces as
\begin{eqnarray}
\delta_{\rm eu}=-\frac{3y^2-2y+4}{2y^2+8}t_1.
\label{bph}
\end{eqnarray}
Here the restriction of the metric is chosen in such a manner that only when $B_0$ is non-zero, oscillation
probability is significant. That is possible only when metric deviates from spherical symmetry. However,
it can be easily checked that even in the case of spherically symmetric metric, say for Robertson-Walker universe, 
non-zero probability of neutrino oscillation can be achieved which may depend on $B_1,B_2,B_3$. Certainly, $B_1,B_2,B_3$
have to have a significant value to visualize the non-zero oscillation phase.  
 
Next example can be given for a black hole space-time when the curvature effect is significant. It can be noted
that in other context the neutrino oscillations around Kerr black holes (in case of active galactic nuclei) were studied
\cite{wudka} earlier. Here I consider the rotating black hole space-time, namely Kerr space-time, given as \cite{doran}
\begin{equation}
d s^2 = \eta_{ij} \, d x^i \, d x^j - \bigg[ \frac{2 \alpha}{\rho} \, s_i \, v_j + \alpha^2 \, v_i \, v_j \bigg] d x^i \, d x^j \label{metbh}
\end{equation}
where
\begin{eqnarray}
\nonumber
\alpha = \frac{\sqrt{2 M r}}{\rho}, ~~~~ ~~~~ \rho^2 = r^2 + \frac{a^2 z^2}{r^2}, \\
s_i = \left(0, ~~ \frac{r x}{\sqrt{r^2 + a^2}}, ~~ \frac{r y}{\sqrt{r^2 + a^2}}, ~~ \frac{z \sqrt{r^2 + a^2}}{r} \right),\\
\nonumber
v_i = \left(1, ~~ \frac{a y}{a^2 + r^2}, ~~ \frac{- a x}{a^2 + r^2}, ~~ 0 \right).
\end{eqnarray}
Here $a$ and $M$ are respectively the specific angular momentum and mass of the Kerr
black hole and $r$ is positive definite satisfying the following equation,
\begin{equation}
r^4 - r^2 \, \left(x^2 + y^2 + z^2 - a^2 \right) - a^2 z^2 = 0.
\end{equation}
                                                                                                                                        
The corresponding non-vanishing component of tetrads (vierbiens) are \cite{doran}
\begin{eqnarray}
\nonumber
&&e^0_t = 1, ~~ ~~  e^1_t = - \frac{\alpha}{\rho} \, s_1, ~~ ~~ e^2_t = - \frac{\alpha}{\rho} \, s_2, ~~ ~~ e^3_t = - \frac{\alpha}{\rho} \, s_3, \label{tet1}  \\
\nonumber
&&e^1_x = 1 - \frac{\alpha}{\rho} \, s_1 \, v_1, ~~ ~~ e^2_x = - \frac{\alpha}{\rho} s_2 \, v_1, ~~ ~~ e^3_x = - \frac{\alpha}{\rho} \,
s_3 \, v_1, \label{tet2} \\
&&e^1_y = - \frac{\alpha}{\rho} \, s_1 v_2, ~~ ~~ e^2_y = 1 - \frac{\alpha}{\rho} \, s_2 \, v_2, ~~ ~~ e^3_y = - \frac{\alpha}{\rho} \,
s_3 \, v_2, ~~ ~~ e^3_z = 1 - \frac{\alpha}{\rho} \, s_3 \, v_3 \label{tet3}.
\label{tetnonv}
\end{eqnarray}
                                                                                                                                        
Using (\ref{bd}), (\ref{metbh}) and (\ref{tetnonv}), the
gravitational scalar potential can be evaluated as
\begin{eqnarray}
 B^0 &=& e_{1\lambda}  \left( \partial_3 e_2^\lambda - \partial_2 e_3^\lambda
 \right) + e_{2 \lambda} \left( \partial_1 e_3^\lambda - \partial_3 e_1^\lambda
 \right) + e_{3 \lambda} \left( \partial_2 e_1^\lambda - \partial_1 e_2^\lambda
 \right).
\label{b0}
\end{eqnarray}
Similarly, one can calculate $B^1,B^2,B^3$. As before, if I choose a special case of rotating black hole
such that $B_0p^0>>{\vec B}.{\vec p}$, then again the neutrino oscillation phase will depend only
on $B_0$ which can be evaluated from (\ref{b0}) using (\ref{metbh}). One also can evaluate the oscillation
phase for a non-rotating (Schwarzschild) black hole. In that case $B^0$ will be zero and thus to get a significant
oscillation, above restriction to the space-time ($B_0p^0>>{\vec B}.{\vec p}$) must be removed and the
oscillation may depend upon $B_1,B_2,B_3$.
   
From the above discussion, it is clear that neutrino oscillations may occur in curved back-ground. For examples
I have chosen a space-time in early universe as well as around a black hole. It is to be noted that
this oscillation phase is independent of the rest mass of eigenstates, and it arises solely due to the effect
of space-time curvature. The amplitude of oscillation is zero when $\theta=0,\pi/2$ and maximum for 
$\theta=\pi/4$. Also from (\ref{pab}) and (\ref{phf}) one can evaluate the oscillation length, $L_{osc}$, as
\begin{eqnarray}
L_{osc}\sim t_1=\frac{\pi}{B_0}
\label{ol}
\end{eqnarray}
(keeping in mind our chosen system of unit as $c=1$ and then $L_{osc}\sim c\,t_1$), when the neutrino is chosen to be moving 
in a speed close to that of light. Now if I put the dimensions appropriately, (\ref{ol}) 
reduces as
\begin{eqnarray}
L_{osc}\sim c\,t_1=\frac{\pi\,\ch}{B_0}.
\label{old}
\end{eqnarray}
If one considers the neutrino$-$anti-neutrino oscillation in the atmosphere of earth where the curvature
effect, $B_0\sim 10^{-40}$ erg, can be found on a satellite orbiting earth with 
velocity, $v\sim 1$ km/sec \cite{mmp}, then the oscillation length, $L_{osc}$, at that
space comes out to be of the order of $10^{8}$ km.

The interesting point can be noted again that the calculated oscillation length is independent
of the rest mass of the individual eigenstates. Whether a mass difference exists or not
oscillation exists, and in a finite
interval a neutrino can retain back to its initial state due to the effect of space-time 
curvature.

\vskip1cm
\noindent{\large Acknowledgment}\\
I thank D. V. Ahluwalia and Sandhya Choubey for their illuminating discussion regarding various important issues,
bringing my attention to some of current results and careful reading the manuscript
which finally have brought the paper in its present form. 


\end{document}